\documentclass[conference,10pt]{IEEEtran}
\IEEEoverridecommandlockouts
\usepackage{algorithm}
\usepackage{cite}
\usepackage{amsmath,amssymb,amsfonts,mathtools}
\usepackage{algorithmic}
\usepackage{graphicx}
\usepackage{textcomp}
\usepackage{xcolor}
\usepackage{braket}
\usepackage{booktabs}
\usepackage{enumitem}
\usepackage{multirow}
\usepackage{hyperref}
\usepackage{url}
\IfFileExists{bbm.sty}{\usepackage{bbm}}{%
  \providecommand{\mathbbm}[1]{\mathbb{#1}}}

\DeclareMathOperator{\sgn}{sgn}

\def\BibTeX{{\rm B\kern-.05em{\sc i\kern-.025em b}\kern-.08em
    T\kern-.1667em\lower.7ex\hbox{E}\kern-.125emX}}

\begin{document}

\title{Depth-Efficient Quantum Topological Data Analysis\\for Regime-Specific Detection of Financial Stress}

\author{
\IEEEauthorblockN{Arul Rhik Mazumder}
\IEEEauthorblockA{\textit{Carnegie Mellon University}\\
Pittsburgh, PA, USA\\
\href{https://orcid.org/0000-0002-2395-4400}{ORCID: 0000-0002-2395-4400}}
\and
\IEEEauthorblockN{Shreyan Ronit Mazumder}
\IEEEauthorblockA{\textit{Cambridge Rindge and Latin School}\\
Cambridge, MA, USA}
}

\maketitle

\begin{abstract}
We present, to our knowledge, the first adaptation of Pauli
Correlation Encoding (PCE) to quantum topological data analysis,
reformulating Betti number estimation as a depth-efficient
variational optimization over a compressed qubit register.
From a Takens embedding and Vietoris--Rips filtration of S\&P~500
returns, we extract combinatorial Laplacians and recast null-space
counting as a continuous-PCE Rayleigh-quotient minimization with
variational deflation, encoding $n_k$ simplex indices into
$O(n_k^{1/\kappa})$ qubits with shallow, ancilla-free circuits.
Because the resulting loss is rational rather than bilinear in the
correlators, the barren-plateau bound of~\cite{Sciorilli25} does
not transfer; empirically the gradient variance decays only
polynomially, with no exponential barren plateau, over $n=4$--$12$
qubits. The classical stage matches
\texttt{ripser}~\cite{bauer2021ripser} on all 190 sliding windows
(2007--2009). On the real market Laplacians ($\beta_1=1$--$22$),
PCE-VQE recovers no null vector from a random start, but warm-starting
from a classical null-space surrogate recovers $\beta_1$ exactly on
all seven windows tested ($n_k\le 259$), placing the obstacle in the optimization landscape rather
than the encoding. Chronologically split
classification gives in-regime ROC AUC $0.818$, but
out-of-distribution evaluation on the 2020 COVID shock and 2022
rate cycle (AUC $0.009$, $0.515$) shows the calibration does not
generalize across crisis regimes.
\end{abstract}


\section{Introduction}

Financial markets exhibit nonlinear dynamics that are poorly captured
by standard statistical summaries. The 2008 financial crisis emerged
from correlations and
structural instabilities invisible to conventional volatility measures
such as variance or VaR\@. During stable regimes a time series $\{x_t\}$
exhibits small, roughly Gaussian fluctuations; approaching a crash, the
underlying dynamical attractor undergoes topological deformation that
can precede the dislocation. In particular, Gidea and Katz~\cite{GK18}
showed that persistence-based topological features of a
multi-index point cloud (S\&P 500, DJIA, NASDAQ, and Russell
2000 viewed jointly as coordinates in $\mathbb{R}^4$) exhibit a
sustained rise in the spectral density of $L^p$-norms of persistence
landscapes for roughly 250 trading days prior to the Lehman
bankruptcy. This motivates the search for early warning signals that
operate on the geometry of market data rather than its pointwise
statistics, and in particular on its topology, which is stable under
perturbations and captures qualitative structural change that purely
local statistics miss~\cite{Car09, EH10, Otter17}. Our pipeline is
inspired by this line of work but adopts a different (single-index
Takens delay) embedding, described in
Section~\ref{sec:methodology}.

Topological Data Analysis (TDA) offers a robust framework for extracting
such geometric features. The central invariants are Betti numbers
$\beta_k$, which count the number of $k$-dimensional holes in the data
manifold: $\beta_0$ counts connected components, $\beta_1$ counts loops,
and $\beta_2$ counts voids. However, computing Betti numbers classically
requires constructing and diagonalizing combinatorial Laplacians whose
dimensions grow combinatorially with the number of simplices, and dense
diagonalization costs $O(n_k^3)$ per window.

Quantum algorithms for TDA (qTDA) have been proposed to address this
computational bottleneck, beginning with the LGZ
algorithm~\cite{lloyd2016quantum} and its subsequent
refinements~\cite{Ubaru21, McArdle22, Gyurik22, Hayakawa22, Akhalwaya22,
Berry24}. These approaches use quantum phase estimation (QPE) to
estimate the eigenspectrum of the combinatorial Laplacian, from which
the Betti number is recovered as the nullity. However, standard QPE
requires deep circuits with many ancilla qubits, making it impractical
for near-term and early fault tolerant hardware. Moreover, recent
complexity-theoretic analyses~\cite{CrichignoKohler24, Schmidhuber23, Berry24} have shown that
superpolynomial quantum advantage requires exponentially large Betti
numbers and specific spectral gap conditions, raising the bar for
practical near-term quantum TDA.

In parallel, Pauli Correlation Encoding (PCE)~\cite{Sciorilli25} has
emerged as a qubit-efficient variational framework for combinatorial
optimization, encoding $m = O(n^k)$ binary variables into $n$ qubits via
$k$-body Pauli correlators. PCE has demonstrated competitive performance
on MaxCut~\cite{Sciorilli25}, portfolio
optimization~\cite{SolovievKrompiec25}, and the LABS
problem~\cite{Sciorilli25LABS}, with provable super-polynomial barren
plateau mitigation for losses of a specific bilinear form in
Pauli expectations, and shallow circuit depths scaling sublinearly in
the number of variables.

\paragraph*{Scope.}

\begin{figure}[t]
  \centering
  \includegraphics[width=\columnwidth]{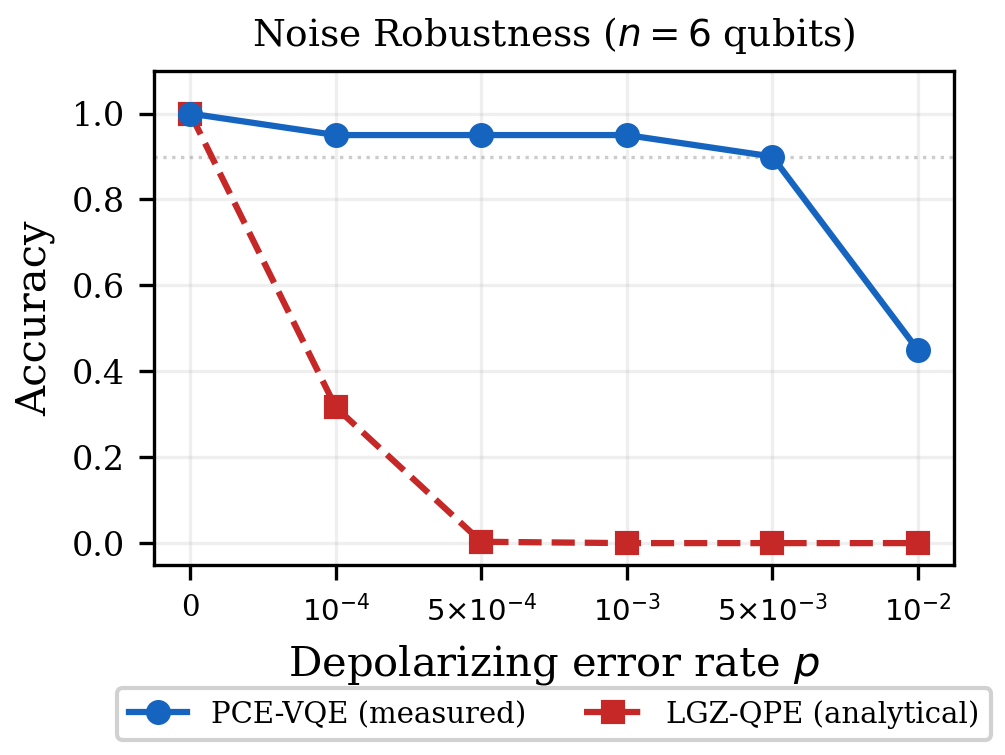}
  \caption{Noise robustness (illustrative; not a matched-scale
  comparison): PCE-VQE accuracy (measured, blue, $n=6$ qubits,
  $n_k \approx 63$ under $\kappa=2$) vs.\ LGZ-QPE at $n_k = 16$
  (measured, red solid) and analytical extrapolation (red dashed).
  PCE maintains ${>}90\%$ accuracy up to $p{=}5{\times}10^{-3}$;
  simulated LGZ-QPE collapses above $p{=}10^{-4}$, matching the
  analytical bound. The two curves are not at matched encoded
  problem size: PCE encodes $n_k\approx 63$ basis dimensions while
  LGZ operates on $n_k=16$. The figure illustrates the depth/noise
  trade-off at moderate scale rather than a like-for-like comparison;
  see Section~\ref{sec:noise_results} for explicit caveats.}
  \label{fig:noise_robustness}
\end{figure}

The classical TDA pipeline (CE benchmark) agrees with
\texttt{ripser} on all 190 sliding windows. PCE-VQE recovers Betti
numbers on six toy Laplacians ($\beta_1 \leq 2$) and on a $\beta_1=4$
four-cycle benchmark under tuned parameters, and its gradient variance
decays only polynomially over $n=4$--$12$ qubits. On the real market
Laplacians ($\beta_1=1$--$22$, $n_k\in[31,429]$) it recovers no null
vector from a random start. Warm-started from the classical null space
it recovers $\beta_1$ exactly (Sec.~\ref{sec:realwindow}), but that
run is a classical-quantum hybrid and does not determine $\beta_1$
independently. The cost crossover at $n_k\gtrsim 10^4$ is an
extrapolation, the noise comparison (Fig.~\ref{fig:noise_robustness})
is not at matched encoded problem size, and the $\beta_1$ classifier
does not generalize across crisis regimes.

In this work, we bridge the qTDA and PCE directions by adapting the
PCE framework to the spectral problem of Betti number estimation.
Two adaptations distinguish our approach from canonical PCE: (i)~we
use the continuous expectation values $c_i =
\langle\Pi_i\rangle\in[-1,+1]$ as eigenvector coefficients, rather
than the binary $x_i = \sgn(\langle\Pi_i\rangle)$ outputs that PCE
was designed to produce; and (ii)~the resulting Rayleigh-quotient
loss with variational deflation is rational in the correlators, so
the formal trainability proof of~\cite{Sciorilli25} does not directly
transfer. Our contributions are: (1)~a complete qTDA pipeline
blueprint from raw financial time series to Betti number estimation;
(2)~to our knowledge, the first application of PCE to eigenvalue
counting, which recasts Betti number estimation as a variational
optimization over $O(n_k^{1/\kappa})$ qubits with depth polynomial in
$n_k^{1/\kappa}$, trading the logarithmic qubit count of LGZ for
shallower, ancilla-free circuits; (3)~a continuous-PCE deflation
protocol with explicit discussion of why the formal barren-plateau
guarantee of~\cite{Sciorilli25} does not transfer, and an empirical
finite-size gradient-variance study; and (4)~numerical experiments
covering classical CE/\texttt{ripser} validation on real S\&P~500
data, PCE-VQE validation on toy Laplacians, chronologically split
classification with independent OOD evaluation, $\kappa=2$ vs.\
$\kappa=3$ resource comparison, and matched-scale noise simulation
at $n_k\leq 16$.
For reproducibility, the full code, figures, and result artifacts used
throughout this paper are publicly available at
\url{https://github.com/arulrhikm/Quantum-Market-Crash-TDA}.

The remainder of this paper: Sec.~\ref{sec:background} introduces
background; Sec.~\ref{sec:related} surveys related work;
Sec.~\ref{sec:methodology} details the four-stage pipeline;
Sec.~\ref{sec:ce} presents the classical benchmark;
Sec.~\ref{sec:setup}--\ref{sec:results} cover experimental setup and
results; Sec.~\ref{sec:conclusion} concludes.

\section{Background}\label{sec:background}

\subsection{Topological Data Analysis and Betti Numbers}

TDA provides tools for extracting the shape of data from point clouds
via simplicial complexes~\cite{Car09, EH10, Otter17}. A $k$-simplex is
a collection of $k+1$ vertices with all $\binom{k+1}{2}$ pairwise
edges; a simplicial complex $\mathcal{K}$ is a collection of simplices
closed under taking faces. Given a point cloud $\{z_i\}_{i=1}^n$ in
$\mathbb{R}^m$ with distance function $d$, the Vietoris--Rips complex
at scale $\varepsilon$ includes a simplex
$\sigma = \{z_{i_0}, \ldots, z_{i_k}\}$ whenever $\|z_{i_a} -
z_{i_b}\| \leq \varepsilon$ for all pairs $a, b$. Varying $\varepsilon$
yields a filtration of nested complexes whose persistent homology
captures multiscale topological structure~\cite{ZC05, Otter17}.

The boundary operator $\partial_k : C_k \to C_{k-1}$ maps $k$-chains to
$(k{-}1)$-chains. The $k$-th combinatorial (Hodge) Laplacian is defined
as
\begin{equation}
    \Delta_k \;=\; \partial_{k+1}\partial_{k+1}^\top
                 + \partial_k^\top \partial_k
    \;\in\; \mathbb{R}^{n_k \times n_k},
    \label{eq:laplacian}
\end{equation}
where $n_k = |\mathcal{C}_k|$ is the number of $k$-simplices. By the
Hodge Decomposition Theorem, $H_k(\mathcal{K}) \cong \ker(\Delta_k)$,
so
\begin{equation}
    \beta_k = \dim \ker(\Delta_k),
    \label{eq:betti}
\end{equation}
i.e., $\beta_k$ equals the number of zero eigenvalues of
$\Delta_k$~\cite{EH10, ZC05}.

\subsection{Quantum Phase Estimation for TDA}\label{sec:qpe_background}

The LGZ algorithm~\cite{lloyd2016quantum} and its variants estimate
Betti numbers by applying QPE to the unitary $U = e^{i\Delta_k}$. Since
zero eigenvalues of $\Delta_k$ map to eigenvalue $e^{i \cdot 0} = 1$ of
$U$, the probability of measuring phase $\theta = 0$ in the QPE
register, when starting from a maximally mixed state,
yields~\cite{Khandelwal23}:
\begin{equation}
    \tilde{\beta}_k = 2^q \cdot p(0),
    \label{eq:betti_est}
\end{equation}
where $q = \lceil \log_2 n_k \rceil$ is the number of system qubits.
The combinatorial Laplacian must be padded to dimension $2^q$;
following~\cite{Khandelwal23}, we use identity padding with
$\tilde{\lambda}_{\max}/2$ on the diagonal to avoid introducing
spurious zero eigenvalues. Standard QPE requires $O(q)$ ancilla qubits
and circuit depths scaling with the desired precision, which limits
near-term applicability.

\subsection{Time-Delay Embedding}\label{sec:takens}

Given a scalar time series $X = (x_0, \ldots, x_{L-1})$, Takens'
embedding~\cite{Tak81} forms delay vectors
\begin{equation}
    \mathbf{z}_t = (x_t,\; x_{t+\tau},\; \ldots,\;
                    x_{t+(m-1)\tau}) \in \mathbb{R}^m
    \label{eq:takens}
\end{equation}
for embedding dimension $m$ and delay $\tau$. By Takens' theorem, for
generic smooth dynamical systems on a $d$-dimensional attractor, this
map is a diffeomorphism when $m \geq 2d+1$, preserving the topology of
the attractor. The delay $\tau$ is selected as the first local minimum
of the mutual information $I(x_t;\,x_{t+\tau})$, and $m$ as the
smallest dimension for which the false nearest neighbor fraction falls
below~5\%~\cite{KS04}; the concrete application of these criteria to
our S\&P~500 series is described in Section~\ref{sec:step1}.

\subsection{Pauli Correlation Encoding}\label{sec:pce_background}

Pauli Correlation Encoding (PCE)~\cite{Sciorilli25} is a
qubit-efficient variational framework that encodes $m = O(n^k)$
classical binary variables into $n$ qubits via $k$-body Pauli
correlators. Given an $n$-qubit parameterized state
$|\Psi(\vec{\theta})\rangle$, a set of $m$ Pauli strings $\Pi^{(k)} =
\{\Pi_1^{(k)}, \ldots, \Pi_m^{(k)}\}$ is chosen, where each
$\Pi_i^{(k)}$ is a permutation of $P_1^{\otimes k} \otimes
\mathbbm{1}^{\otimes (n-k)}$ for Pauli matrices $P_1 \in \{X, Y, Z\}$.
The $i$-th binary variable is then decoded via
\begin{equation}
    x_i = \sgn\!\left(
        \langle \Psi(\vec{\theta}) |
        \Pi_i^{(k)} |
        \Psi(\vec{\theta}) \rangle
    \right).
    \label{eq:pce_decode}
\end{equation}
Our instantiation follows the standard, reproducible enumeration
strategy used in introductory PCE implementations~\cite{Sciorilli25,
IBMpce}: enumerate $\kappa$-local positions and operator assignments
from $\{X,Y,Z\}^\kappa$, place identities elsewhere, and truncate to
the required count. More sophisticated, topology-aware assignments and
commuting-group measurement reductions are compatible with PCE but are
left to future work in this application. For $k=2$ (quadratic PCE),
this yields $m = O(n^2)$ variables from $n$ qubits; for $k=3$ (cubic
PCE), $m = O(n^3)$. A hardware-efficient ansatz (HEA) with alternating
single-qubit rotation and entangling layers is trained to minimize a
loss function $\mathcal{L}(\vec{\theta})$ constructed from the Pauli
expectations.

The key properties of PCE that motivate our adaptation are:
\begin{enumerate}
    \item \textbf{Polynomial qubit compression:} $m$ problem variables
    are encoded in $O(m^{1/k})$ qubits, yielding quadratic or cubic
    space savings.
    \item \textbf{Built-in barren plateau mitigation:} Sciorilli et al.~\cite{Sciorilli25} prove
    that the variance of the gradient of cost functions of the form
    $\mathcal{L} = \sum_{ij} w_{ij}\langle\Pi_i\rangle\langle\Pi_j
    \rangle$ is lower-bounded by a super-polynomial function
    of $n$, avoiding the exponential gradient vanishing of generic
    HEA+global cost function combinations~\cite{Cerezo21}. Note that this
    guarantee applies to the canonical bilinear PCE loss; the
    Rayleigh-quotient loss with deflation introduced in this paper is
    rational in the correlators and is not directly covered by
    the proof (Section~\ref{sec:deflation}).
    \item \textbf{Shallow circuits:} Circuit depth scales sublinearly
    in $m$, and the number of parameters scales linearly.
    \item \textbf{Commuting measurement groups:} The Pauli set
    $\Pi^{(k)}$ decomposes into a small number of mutually commuting
    subsets, enabling efficient simultaneous measurement.
\end{enumerate}

\paragraph*{PCE experiments in literature.} The Sciorilli et al.\ Nature Communications study~\cite{Sciorilli25}
reports MaxCut numerical simulations at $m = 7000$ and trapped-ion
experiments on IonQ Aria-1 / Quantinuum H1-1 with $n=17$ qubits and
unweighted MaxCut sizes $m \in \{800, 2000\}$ (weighted $m=512$). The
follow-on PCE-LABS paper~\cite{Sciorilli25LABS} operates in a
different regime: simulations of LABS instances up to $N=44$--$45$
with $n=4$--$6$ qubits and $\sim 30$ two-qubit gates, with
proof-of-principle experiments on IonQ Forte. The two studies are
complementary, not directly comparable in scale; we attribute
MaxCut-scale numbers to~\cite{Sciorilli25} and LABS-scale numbers
to~\cite{Sciorilli25LABS}.

\section{Related Work}\label{sec:related}

\subsection{Quantum Algorithms for TDA}

The landscape of quantum TDA algorithms has evolved rapidly since the
original LGZ proposal~\cite{lloyd2016quantum}.

\textbf{LGZ and refinements.} The original algorithm uses QPE on
$e^{i\Delta_k}$ prepared from a maximally mixed state to estimate
$\beta_k$ via~\eqref{eq:betti_est}. Ubaru et al.~\cite{Ubaru21}
improved the exponential speedup and depth complexity. McArdle et
al.~\cite{McArdle22} proposed a streamlined algorithm achieving an
almost quintic speedup over rigorous classical methods using
exponentially fewer qubits, but also introduced a quantum-inspired
classical power method with scaling only quadratically worse, casting
doubt on exponential advantage for practical instances.

\textbf{Persistent Betti numbers.} Hayakawa~\cite{Hayakawa22} gave the
first quantum algorithm for persistent Betti numbers of arbitrary
dimensions, efficient for Vietoris--Rips complexes but limited to
normalized estimates.

\textbf{Complexity-theoretic limits.} Crichigno and
Kohler~\cite{CrichignoKohler24} proved that deciding whether a clique
complex has a $k$-dimensional hole is QMA$_1$-hard and that the exact
counting version is \#BQP-hard. Schmidhuber and Lloyd~\cite{Schmidhuber23}
separately showed that approximating Betti numbers to multiplicative
error is NP-hard for general complexes.
Berry et al.~\cite{Berry24} showed that superquadratic quantum
speedups require multiplicative error approximation with asymptotically
growing Betti numbers, and proposed a dequantization showing that
exponentially large dimension and Betti number are necessary but
insufficient for superpolynomial advantage. Their fault-tolerant
resource estimates require tens of billions of Toffoli gates.

\textbf{Near-term approaches.} Scali et al.~\cite{Scali24} proposed a
thermal-state approach (``thermal QTDA'') that reduces Betti number
estimation to a purity test on a low-temperature Gibbs state of
$\Delta_k$, using $2\lceil\log_2 n_k\rceil$ system qubits plus a
SWAP-test ancilla and inheriting an inverse spectral-gap dependence in
the state-preparation time. Relative to our PCE approach it uses
exponentially fewer qubits ($O(\log n_k)$ vs.\ $O(n_k^{1/\kappa})$) but
needs ancilla overhead and Lindbladian machinery; PCE offers shallower,
ancilla-free circuits at the cost of a larger register and no formal
trainability guarantee. Akhalwaya et al.~\cite{Akhalwaya22} compared
quantum and classical Monte Carlo estimators for clique-complex Betti
numbers.

\textbf{Hybrid approaches.} Nghiem and Wei~\cite{NghiemLeeWei25}
proposed a hybrid quantum-classical algorithm combining classical
simplex enumeration with quantum Betti number estimation, using
$(|S_r| + |S_{r+1}|) \cdot O(r)$ total qubits, linear in the simplex count and so many more than
LGZ or our approach. Their
framework provides provable complexity guarantees for normalized Betti
number estimation, whereas our PCE pipeline is a trainable heuristic
targeting absolute Betti numbers without such guarantees.
Liu~\cite{Liu25} proposed bridging from Betti numbers to persistence
diagrams via quantum kernel methods.

\subsection{Pauli Correlation Encoding}

Beyond the foundational MaxCut work~\cite{Sciorilli25}, PCE has been
extended to portfolio optimization by Soloviev and
Krompiec~\cite{SolovievKrompiec25}, the LABS problem by Sciorilli et
al.~\cite{Sciorilli25LABS}, and budget-constrained optimization by
Pad\'in-Mart\'inez et al.~\cite{PadinMartinez26}; IBM Quantum has
published a tutorial implementation~\cite{IBMpce}. All prior PCE
applications target combinatorial optimization, finding a single
optimal bitstring via the binary decoding rule
in~\eqref{eq:pce_decode}. Our work is the first, to our knowledge, to
apply PCE to an eigenvalue counting problem, where the goal is to
determine the multiplicity of the zero eigenvalue of a matrix. To do
so we depart from canonical PCE in two respects: (i)~we use the continuous expectation values
$c_i=\langle\Pi_i\rangle\in[-1,+1]$ as eigenvector coefficients
without applying the $\sgn$ rounding, and (ii)~we introduce a
Rayleigh-quotient loss with variational deflation. Both adaptations
move the cost function outside the bilinear-in-correlators class
covered by the trainability proof of~\cite{Sciorilli25}.

\subsection{Classical TDA for Financial Applications}

TDA has been applied to financial market analysis by several
groups~\cite{GK18, Otter17}. Gidea and Katz~\cite{GK18} demonstrated
that persistent homology features of a multi-index point cloud
constructed from the daily log returns of four major US stock indices
(S\&P~500, DJIA, NASDAQ, and Russell~2000), each used as a coordinate
in $\mathbb{R}^4$, exhibit a sustained rise in the spectral density at
low frequencies of $L^p$-norms of persistence landscapes for roughly
250 trading days prior to the Lehman bankruptcy. Subsequent work has
shown that persistence-based features provide complementary
information to standard volatility measures for regime classification
and crash forecasting. Our pipeline differs from~\cite{GK18} in two
respects: we use a single-index S\&P~500 Takens delay
embedding rather than a multi-index direct embedding, and we replace
the classical persistence-based summary by a quantum-accelerated
Betti number computation.

\section{Methodology}\label{sec:methodology}

\begin{figure}[t]
  \centering
  \includegraphics[width=\columnwidth]{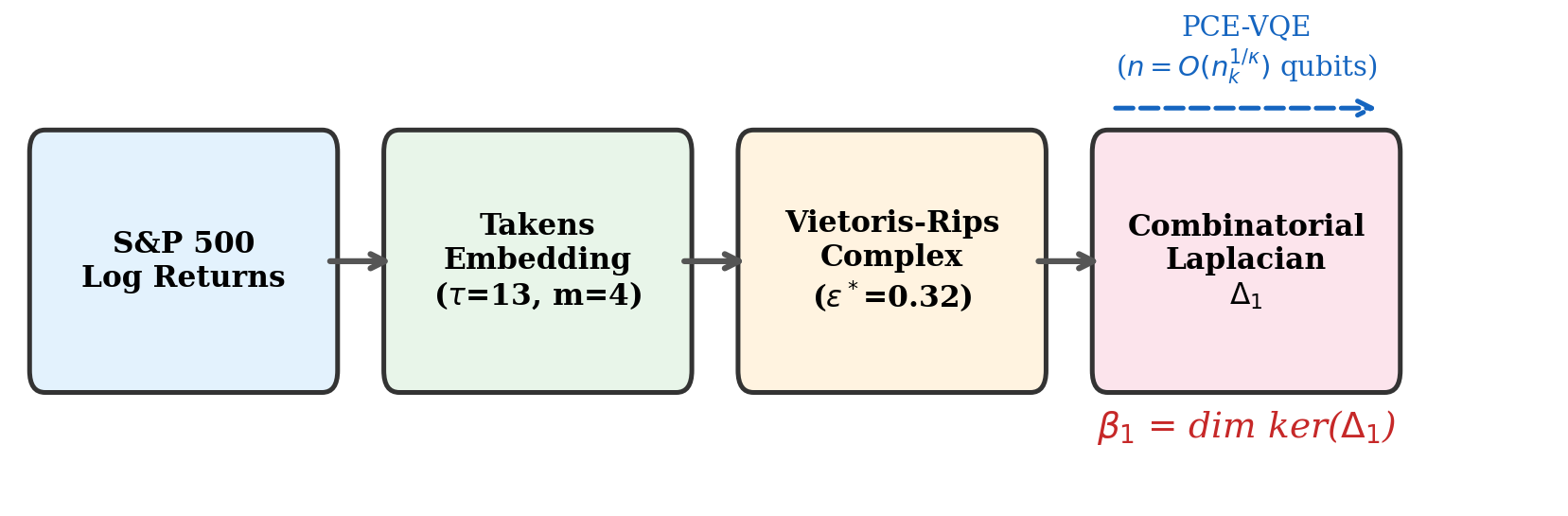}
  \caption{Overview of the qTDA pipeline with PCE qubit compression:
    (1)~S\&P~500 log returns $\to$ Takens embedding ($\tau=13$, $m=4$)
    $\to$ point cloud in $\mathbb{R}^4$;
    (2)~Vietoris--Rips complex $\mathrm{VR}(\varepsilon^*)$, 461 points
    per window;
    (3)~combinatorial Laplacian $\Delta_1 = \partial_2\partial_2^\top +
    \partial_1^\top\partial_1$, $n_k \in [31,\,429]$;
    (4)~PCE encoding into $n = O(n_k^{1/\kappa})$ qubits, HEA + COBYLA
    optimization with variational deflation $\to$ $\tilde{\beta}_1$.}
  \label{fig:pipeline}
\end{figure}

Our pipeline converts raw financial time series into quantum-estimated
Betti numbers through four stages (Fig.~\ref{fig:pipeline}).

\subsection{Step 1: Time-Delay Embedding}\label{sec:step1}

We apply Takens' embedding~\eqref{eq:takens} to daily S\&P~500 log
returns $r_t = \log(P_t / P_{t-1})$, determining $\tau$ and $m$ via
mutual information and false nearest neighbors (FNN) directly on the
return series.

\paragraph*{Selecting $\tau$.}
The delay $\tau$ is the first local minimum of the mutual information
$I(r_t; r_{t+\tau}) = \sum_{x,y} p(x,y)\log[p(x,y)/(p(x)p(y))]$,
evaluated at lags $\tau = 1, \ldots, 30$ via a $16{\times}16$
histogram of the pairs $(r_t, r_{t+\tau})$. Mutual information is
high at small $\tau$ (successive coordinates are redundant) and
decreases as $\tau$ grows; its first local minimum identifies the
lag at which $r_{t+\tau}$ is maximally informative without becoming
noise. For our return series this minimum occurs at
$\tau = 13$ trading days (${\approx}2.5$ weeks), reflecting the
characteristic decorrelation timescale of daily equity returns.

\paragraph*{Selecting $m$.} With $\tau = 13$ fixed, $m$ is the
smallest dimension for which fewer than~5\% of nearest neighbors
are false. For each embedded point $\mathbf{z}_t$, its nearest
neighbor $\mathbf{z}_{t'}$ in the $m$-dimensional embedding is
declared false if the ratio
$R_t = |x_{t+m\tau}-x_{t'+m\tau}| / \|\mathbf{z}_t-\mathbf{z}_{t'}\|_m$
exceeds a threshold (we use $R_{\text{tol}}=15$, within the
conventional $10$--$15$ range~\cite{KennelBrownAbarbanel92});
$x_{t+m\tau}$ is the candidate $(m{+}1)$-th coordinate. The FNN
fraction remains above~5\% at $m = 1, 2, 3$ and drops below~5\% at
$m = 4$, giving the selected embedding dimension. Although Takens'
bound formally requires $m \geq 2d+1$, this bound is conservative
and the attractor unfolds sufficiently for downstream TDA at $m=4$
in practice~\cite{KS04}.

\paragraph*{Resulting point cloud.}
With $\tau = 13$ and $m = 4$, each sliding window of $W = 500$
trading days yields $n_{\mathrm{pts}} = W - (m-1)\tau = 461$
embedded points in $\mathbb{R}^4$.

\subsection{Step 2: Vietoris--Rips Complex Construction}

For each windowed point cloud, we construct the Vietoris--Rips complex
$\mathrm{VR}(\varepsilon)$ at a range of thresholds $\varepsilon$.
Varying $\varepsilon$ produces a filtration of nested complexes;
tracking how topological features appear and disappear across this
filtration yields persistence diagrams, multiscale summaries that are stable
under noise and perturbations~\cite{EH10, ZC05}. Applied to the sliding
window, this procedure produces a time series of Betti curves whose
changes signal regime transitions in the underlying financial
dynamics~\cite{GK18}.

\paragraph*{Choosing the working threshold $\varepsilon^*$.}
For classical validation with \texttt{ripser} we use the full
distance-induced Vietoris--Rips filtration. \texttt{ripser} returns
persistence intervals $\{(b_i, d_i)\}$ for each homological dimension
$k$; we obtain a Betti number at scale $\varepsilon$ by counting
intervals with $b_i \leq \varepsilon < d_i$ (the standard
``slice-of-the-barcode'' definition). For the CE benchmark and
quantum experiments, which operate at a single scale, we select a
working threshold $\varepsilon^*$ by scanning a grid constructed from
percentiles of the interpoint distance distribution and choosing the
$\varepsilon$ that maximizes a robust central tendency (median) of
$\beta_1$ across windows, with a stability tie-break toward smaller
$\varepsilon$. This yields $\varepsilon^* = 0.32$ for the 2007--2010
dataset. This selection uses no crash labels, but because it is made on
windows that overlap the 2007--2010 test split of
Sec.~\ref{sec:classification}, $\varepsilon^*$ is not strictly
out-of-sample for that evaluation.

\subsection{Step 3: PCE Encoding of the Spectral Problem}
\label{sec:pce_encoding}

This section describes our central methodological contribution: the
adaptation of Pauli Correlation Encoding to the problem of estimating
$\beta_k = \dim\ker(\Delta_k)$.

\subsubsection{From eigenvalue counting to variational optimization}

Given the combinatorial Laplacian
$\Delta_k \in \mathbb{R}^{n_k \times n_k}$, we seek the number of
zero eigenvalues. We reformulate this as a sequence of variational
feasibility problems. In standard VQE, one would encode $\Delta_k$ as
a Hamiltonian on $\lceil \log_2 n_k\rceil$ qubits and minimize the
expectation value directly. Here, we instead use the PCE correlators
to parameterize a trial vector in $\mathbb{R}^{n_k}$ and evaluate the
Rayleigh quotient classically (Eq.~\eqref{eq:pce_loss} below). For
conceptual motivation, the standard VQE objective for the first
eigenvalue is
\begin{equation}
    \mathcal{L}_1(\vec{\theta})
    = \bra{\Psi(\vec{\theta})} \Delta_k \ket{\Psi(\vec{\theta})},
    \label{eq:loss_1}
\end{equation}
where $|\Psi(\vec{\theta})\rangle$ lives in the $n_k$-dimensional
simplex space; the PCE adaptation replaces this with a classical
evaluation over the compressed qubit register.

\subsubsection{Continuous-PCE variable encoding}

The Laplacian $\Delta_k$ acts on the vector space $\mathbb{R}^{n_k}$
indexed by $k$-simplices. Rather than encoding this space directly
into $\lceil \log_2 n_k \rceil$ qubits (as in standard LGZ), we use
PCE to encode $n_k$ basis coefficients into $n = O(n_k^{1/\kappa})$
qubits for a chosen compression order $\kappa$.

Concretely, we select a set of $n_k$ Pauli strings $\Pi^{(\kappa)} =
\{\Pi_1, \ldots, \Pi_{n_k}\}$ as described in
Section~\ref{sec:pce_background}. We depart from the canonical PCE
framework~\cite{Sciorilli25} by treating each Pauli expectation as a
continuous eigenvector coefficient,
\begin{equation}
    c_i(\vec{\theta}) =
    \langle \Psi(\vec{\theta}) | \Pi_i | \Psi(\vec{\theta}) \rangle
    \in [-1, +1],
    \quad i = 1, \ldots, n_k,
    \label{eq:pce_coefficients}
\end{equation}
rather than rounding to $\sgn(\langle\Pi_i\rangle)$ as is done in the
canonical optimization setting. The reachable set $\{\mathbf{c}(\vec
\theta)\}$ is therefore a strict subset of $[-1,+1]^{n_k}$, and the
Rayleigh quotient evaluated below operates on this restricted ansatz
rather than on all of $\mathbb{R}^{n_k}$. We discuss the resulting
expressivity question explicitly in Section~\ref{sec:discussion}.

In our experiments, we instantiate $\Pi^{(\kappa)}$ via a simple
deterministic enumeration consistent with the PCE
framework~\cite{Sciorilli25, IBMpce}: for each $\kappa$-tuple of
distinct qubit indices and each operator assignment in
$\{X,Y,Z\}^\kappa$, we place the chosen Paulis on those indices and
identity elsewhere (reversing bit order to match simulator
conventions), appending strings until $|\Pi^{(\kappa)}| = n_k$. We
choose $n$ so that $\binom{n}{\kappa}\cdot 3^\kappa \ge n_k$. This
provides a reproducible mapping from simplex indices to Pauli strings.
We did not optimize this assignment for the Laplacian's topology, nor
apply commuting-group measurement reductions in our implementation;
these are promising directions for future work. The variational loss
function becomes
\begin{equation}
    \mathcal{L}(\vec{\theta})
    = \frac{\mathbf{c}(\vec{\theta})^\top \Delta_k\,
            \mathbf{c}(\vec{\theta})}
           {\|\mathbf{c}(\vec{\theta})\|^2}
    = \frac{\sum_{i,j} [\Delta_k]_{ij}\, c_i\, c_j}
           {\sum_i c_i^2},
    \label{eq:pce_loss}
\end{equation}
where $\mathbf{c}(\vec{\theta}) = (c_1, \ldots, c_{n_k})^\top$. This
is a Rayleigh quotient over the restricted set of vectors
parameterized by PCE correlators. Minimizing~\eqref{eq:pce_loss}
yields the smallest eigenvalue of $\Delta_k$ accessible from
within the reachable set; if this is below threshold $\delta$, an
approximate null vector has been found.

\subsubsection{Measurement protocol}

The loss~\eqref{eq:pce_loss} requires estimating second-order
correlators $c_i\, c_j = \langle \Pi_i \rangle \langle \Pi_j \rangle$.
Since the Pauli set $\Pi^{(\kappa)}$ decomposes into a small number
$G$ of mutually commuting groups (typically $G = 3$ for $\kappa =
2$~\cite{Sciorilli25}), all $n_k$ expectations can be estimated from
$G$ measurement bases per optimization step. The total per-step
measurement overhead is $O(G / \epsilon_{\mathrm{shot}}^2)$ shots,
independent of $n_k$.

We note that the Vietoris--Rips Laplacian may admit additional
commutation structure from the simplicial complex topology,
potentially allowing further grouping reduction beyond generic PCE; we
leave the derivation of tighter bounds on $G$ for structured
Laplacians to future work.

\subsubsection{Evaluating the Rayleigh quotient}

Evaluating~\eqref{eq:pce_loss} requires computing
$\sum_{ij} [\Delta_k]_{ij}\, c_i\, c_j$ from the measured expectations
$\{c_i\}$. Since $\Delta_k$ is sparse (each row has at most
$O(n_{\mathrm{pts}})$ nonzero entries for the VR complex, though
empirically far fewer), this sum is evaluated classically in the
optimization loop. Across all 190 windows the mean number of nonzero
entries per row of $\Delta_1$ is 3.48 (window-averaged maximum 8.8),
confirming that the classical post-processing cost per optimization
step is negligible compared to the quantum measurement cost.

\subsection{Step 4: Variational Deflation for Betti Number Counting}
\label{sec:deflation}

Estimating $\beta_k$ requires counting the multiplicity of the
zero eigenvalue, not just finding a single ground state. We employ
variational deflation~\cite{Higgott19}: after finding the $j$-th
approximate null vector $\mathbf{c}^{(j)}$, we add a penalty term to
the loss for subsequent searches:
\begin{equation}
    \mathcal{L}_{j+1}(\vec{\theta})
    = \mathcal{L}(\vec{\theta})
      + \mu \sum_{\ell=1}^{j}
        \left|
        \frac{\mathbf{c}(\vec{\theta})^\top
              \mathbf{c}^{(\ell)}}
             {\|\mathbf{c}(\vec{\theta})\|
              \|\mathbf{c}^{(\ell)}\|}
        \right|^2,
    \label{eq:deflation}
\end{equation}
where $\mu > 0$ is a penalty strength (default $\mu=5.0$ unless
stated otherwise). This encourages the optimizer to find states
orthogonal to previously identified null vectors. The Betti number is
estimated as the number of successful null-space searches before the
minimum of $\mathcal{L}_{j+1}$ exceeds the threshold $\delta$ (default
$\delta=0.01$):
\begin{equation}
    \tilde{\beta}_k = \max\{j : \mathcal{L}_j^* < \delta\}.
    \label{eq:betti_pce}
\end{equation}

This deflation scheme carries an intrinsic cost. Each null vector is
recovered by its own optimization, so counting the $\beta_k$ zero
eigenvalues requires $\beta_k$ sequential PCE-VQE runs, and the total
circuit-optimization budget grows linearly with the Betti number being
estimated. This is a real weakness in exactly the high-$\beta_k$ regime
that motivates quantum TDA~\cite{Berry24}, and in practice it limits
the multiplicity the scheme can reach. Subspace-search variants that
recover several null vectors in a single optimization
(Sec.~\ref{sec:conclusion}) are the natural route to removing this
linear factor.

\subsubsection{Why the Sciorilli barren-plateau guarantee does not
directly transfer}

The super-polynomial barren plateau lower bound proved
in~\cite{Sciorilli25} applies specifically to loss functions of the
bilinear form $\mathcal{L}_{\mathrm{PCE}} = \sum_{ij} w_{ij}\,
\langle \Pi_i \rangle \langle \Pi_j \rangle$ with bounded coefficient
weights. Our loss~\eqref{eq:pce_loss} has two structural differences
that move it outside the class covered by that proof:
\begin{enumerate}
\item the Rayleigh-quotient denominator $\sum_i c_i^2$
introduces a normalization that makes the loss rational,
rather than bilinear, in the correlators;
\item the deflation penalty in~\eqref{eq:deflation} contributes
additional ratio terms whose Hessian structure is not captured by the
bilinear analysis.
\end{enumerate}
We therefore make no formal trainability claim. Instead, we report
empirical gradient-variance behavior over $n=4$--$12$ qubits and
$j=0,1,2$ deflation rounds in Section~\ref{sec:bp_results}. Across
the tested range the gradient variance decays only polynomially with
system size (log--log slope $\approx -1$), far from the exponential
vanishing of a barren plateau and consistent with the trainability
advantages of PCE-style encodings, but this should not be read as a
verification of the asymptotic Sciorilli bound.

\section{Classical Exact-Eigensolver Benchmark}\label{sec:ce}

To contextualize the quantum pipeline, we introduce a classical
exact-eigensolver (CE) benchmark that computes Betti numbers by
direct dense diagonalization of $\Delta_k$. The CE benchmark serves
two roles: it provides exact ground-truth Betti curves free of the
approximation errors of QPE/VQE, and it establishes the cost regime
at which quantum methods become advantageous.

\subsection{Algorithm and Threshold}

Given $\Delta_k\in\mathbb{R}^{n_k\times n_k}$, the CE benchmark
computes the full eigendecomposition $\Delta_k = Q\Lambda Q^\top$ via
\texttt{numpy.linalg.eigh} (LAPACK \texttt{dsyevd}). The Betti
number is recovered as the nullity
\begin{equation}
  \beta_k^{(\mathrm{CE})} = |\{i:\lambda_i<\delta\}|,\quad
  \delta = c_\delta\cdot\varepsilon_{\mathrm{mach}}\cdot n_k
           \cdot\|\Delta_k\|_2,
  \label{eq:ce_betti}
\end{equation}
with $c_\delta=10$ and $\varepsilon_{\mathrm{mach}}\approx 2.2\times
10^{-16}$. Across all 190 sliding windows we observe a spectral gap
of at least three orders of magnitude between true-zero and nonzero
eigenvalues; the empirical gap distribution is included in the
released code.

\subsection{Complexity and Implementation}

The dominant cost is dense eigendecomposition,
$O(n_k^3)$ time / $O(n_k^2)$ memory. For our pipeline ($W=500$,
$m=4$) the typical Laplacian dimension grows as
$n_1 = \binom{n_{\mathrm{pts}}}{2}\rho(\varepsilon)$. At
$\varepsilon^*$ we find $n_1\in[31,429]$ (mean 136, median 116),
giving eigendecomposition times of $\approx 41$--$228$~ms per window
(mean $\approx 89$~ms) on a single CPU; total runtime for all 190
windows is $\approx 25.5$~s. Across the 190 windows the mean number
of nonzero entries per row of $\Delta_1$ is 3.48 (window-averaged
maximum 8.8), confirming that classical post-processing of the
Rayleigh quotient~\eqref{eq:pce_loss} is negligible relative to the
quantum measurement cost. Mean number of 2-simplices per window is
60 (max 476); edge sparsity relative to the complete graph on 461
points is~0.13\%, reflecting the conservative $\varepsilon^*$ choice
and the principal reason $\beta_1$ stays moderate (mean 4.72).

\subsection{Validation Against \texttt{ripser}}

We verify that the CE benchmark agrees with
\texttt{ripser}~\cite{bauer2021ripser} on all 190 sliding windows in
the 2007--2009 dataset. \texttt{ripser} returns persistence intervals
for $H_0$ and $H_1$; we extract Betti numbers at scale $\varepsilon^*$
by counting intervals $(b_i, d_i)$ with $b_i \leq \varepsilon^* < d_i$,
and compare against the nullity computed by~\eqref{eq:ce_betti}.
Table~\ref{tab:ce_validation} reports the comparison for $\beta_0$
and $\beta_1$ at the selected filtration scale $\varepsilon^*=0.32$.
We emphasize that this validation concerns only the classical Betti extractor used as ground-truth in this paper; the PCE-VQE
quantum pipeline is validated separately in
Section~\ref{sec:pce_convergence} on toy Laplacians.

\begin{table}[t]
\caption{CE vs.\ \texttt{ripser} Agreement
         (2007--2009, $W=500$, $\varepsilon^*=0.32$)}
\label{tab:ce_validation}
\centering
\resizebox{\columnwidth}{!}{%
\begin{tabular}{lcccc}
\toprule
Window period
  & $\beta_0^{\texttt{rip}}$ & $\beta_0^{\mathrm{CE}}$
  & $\beta_1^{\texttt{rip}}$ & $\beta_1^{\mathrm{CE}}$ \\
\midrule
Jan 2007 -- Dec 2008        & 284 & 284 & 15 & 15 \\
Mar 2007 -- Feb 2009        & 296 & 296 &  9 &  9 \\
Jun 2007 -- May 2009        & 327 & 327 &  6 &  6 \\
Sep 2007 -- Aug 2009        & 335 & 335 &  3 &  3 \\
Dec 2007 -- Nov 2009        & 338 & 338 &  3 &  3 \\
\midrule
\textbf{All 190 windows}
  & \textbf{100\%} & --- & \textbf{100\%} & --- \\
\bottomrule
\end{tabular}}
\par\smallskip
{\footnotesize Both $\beta_0$ and $\beta_1$ agreement with
\texttt{ripser}: 190/190 (100\%). $\beta_1 \in [0, 22]$ across all
windows (mean $4.72$). \emph{This table reports classical-vs-classical
agreement; PCE-VQE results are reported in
Section~\ref{sec:pce_convergence}.}}
\end{table}

\section{Experimental Setup}\label{sec:setup}

\subsection{Dataset}

We use daily S\&P~500 closing prices downloaded from Yahoo Finance
(ticker: \texttt{\^{}GSPC}) covering January~2,~2003 to
December~30,~2010, encompassing the 2007--2009 financial crisis. The
dataset contains 2,014 trading days. Log returns
$r_t = \log(P_t/P_{t-1})$ yield 2,013 return observations. After
Takens embedding with $\tau=13$ and $m=4$, the embedded point cloud
contains 1,974 vectors in $\mathbb{R}^4$. Sliding windows of $W=500$
trading days are extracted from the raw return series with step size
$8$~days, producing 190 windows; each window of 500 raw days yields
$500-(m{-}1)\tau = 461$ embedded points in $\mathbb{R}^4$. Points are
standardized per-dimension before distance computation. For
out-of-distribution evaluation
(Section~\ref{sec:classification}) we additionally use S\&P~500 daily
data from June~2019--December~2020 (COVID shock) and
January~2022--June~2023 (rate-cycle drawdown); the Takens pipeline is
applied identically. The 2003--2010 dataset is included in the
released code; OOD episodes are downloaded via \texttt{yfinance} with
fixed date ranges.

\subsection{Classical Baselines}

Two classical baselines are employed:
\begin{enumerate}
    \item \texttt{ripser}~\cite{bauer2021ripser}: exact persistence
    diagrams via boundary-matrix reduction (primary ground truth).
    \item CE benchmark (Section~\ref{sec:ce}): exact Betti numbers via
    dense diagonalization (secondary baseline and resource reference).
\end{enumerate}

\subsection{Quantum Simulation}

\subsubsection{Small-scale exact simulation}
The PCE-variational circuits are implemented in Qiskit and simulated
on BlueQubit statevector backends for $n = 4$--$12$ qubits, encoding
Laplacians of dimension $n_k = 16$--$259$ (quadratic PCE, $\kappa =
2$). These simulations
validate the Betti number estimation protocol on synthetic Laplacians.
Sec.~\ref{sec:realwindow} runs seven real-window Laplacians
($n_k=76$--$259$); larger real windows (up to $n_k=429$) have not been
run through the quantum pipeline. All
statevector-only experiments (convergence, encoding-order comparison,
barren-plateau sweep) use BlueQubit; noise experiments use Qiskit Aer
density matrices by construction.

\subsubsection{Resource extrapolation}
For larger scales ($n_k = 10^3$--$10^5$), we report \emph{analytical}
resource estimates: qubit count, circuit depth, gate count, and
measurement shots, extrapolated from the small-scale empirical data
and the known scaling properties of PCE~\cite{Sciorilli25}. These are
extrapolations, not measurements, and should be read as research
targets rather than achieved capabilities.

\subsubsection{Noise simulation}
We simulate depolarizing noise at error rates
$p \in \{0,\, 10^{-4},\, 5 \times 10^{-4},\, 10^{-3},\, 5 \times
10^{-3},\, 10^{-2}\}$ per gate using Qiskit Aer's density-matrix
backend to assess noise robustness. For LGZ-QPE, we report both
matched-scale simulated accuracy at $n_k \in \{8, 16\}$ and analytical
extrapolation for larger $n_k$. The matched-scale comparison and its
limitations are discussed in detail in
Section~\ref{sec:noise_results}.

\subsection{Hardware-Efficient Ansatz}

We employ a hardware-efficient ansatz~\cite{Kandala17} with
alternating layers of single-qubit rotations ($R_Y$, $R_Z$) and
entangling CNOT gates in a linear connectivity topology. The number of
layers is set to $L = \lceil 2n \rceil$, scaling linearly in the
qubit count $n$ rather than in $m = n_k$ (in
contrast,~\cite{Sciorilli25} reports brickwork depths sublinear in
$m$; the two parameterizations are related by $n = O(m^{1/\kappa})$,
so $L = O(n) = O(m^{1/\kappa})$ here is consistent with the
$O(m^{1/\kappa})$ depth class but slightly more conservative). With
$O(n)$ two-qubit depth per layer, this yields total circuit depth
$O(n^2)$, which we use as the proxy
$d_{\mathrm{proxy}} = 2n^2$ in resource comparisons. The classical
optimizer is COBYLA, which typically converges in 80--130 iterations
on the toy and small-scale benchmarks reported here; the hardware
resource estimate in Section~\ref{sec:hardware_estimate} uses 200
iterations as a conservative upper bound.

\subsection{Resource Metrics}

We track the following quantities as functions of the Laplacian
dimension $n_k$ and PCE order $\kappa$:
\begin{itemize}
    \item Number of qubits: $n = O(n_k^{1/\kappa})$
    \item Number of variational parameters:
          $N_{\mathrm{param}} = O(n \cdot L)$
    \item Two-qubit gate count per circuit evaluation
    \item Circuit depth
    \item Number of measurement bases $G$
    \item Measurement shots per optimization step
    \item Number of deflation rounds (equals $\beta_k$)
    \item Total optimizer iterations across all deflation rounds
\end{itemize}
These are compared between PCE-variational (this work), standard
LGZ-QPE, the streamlined algorithm of~\cite{McArdle22}, and the
thermal QTDA approach of~\cite{Scali24}.

\section{Results}\label{sec:results}

\subsection{Betti Curve Comparison (Classical Pipeline)}

\begin{figure}[t]
  \centering
  \includegraphics[width=\columnwidth]{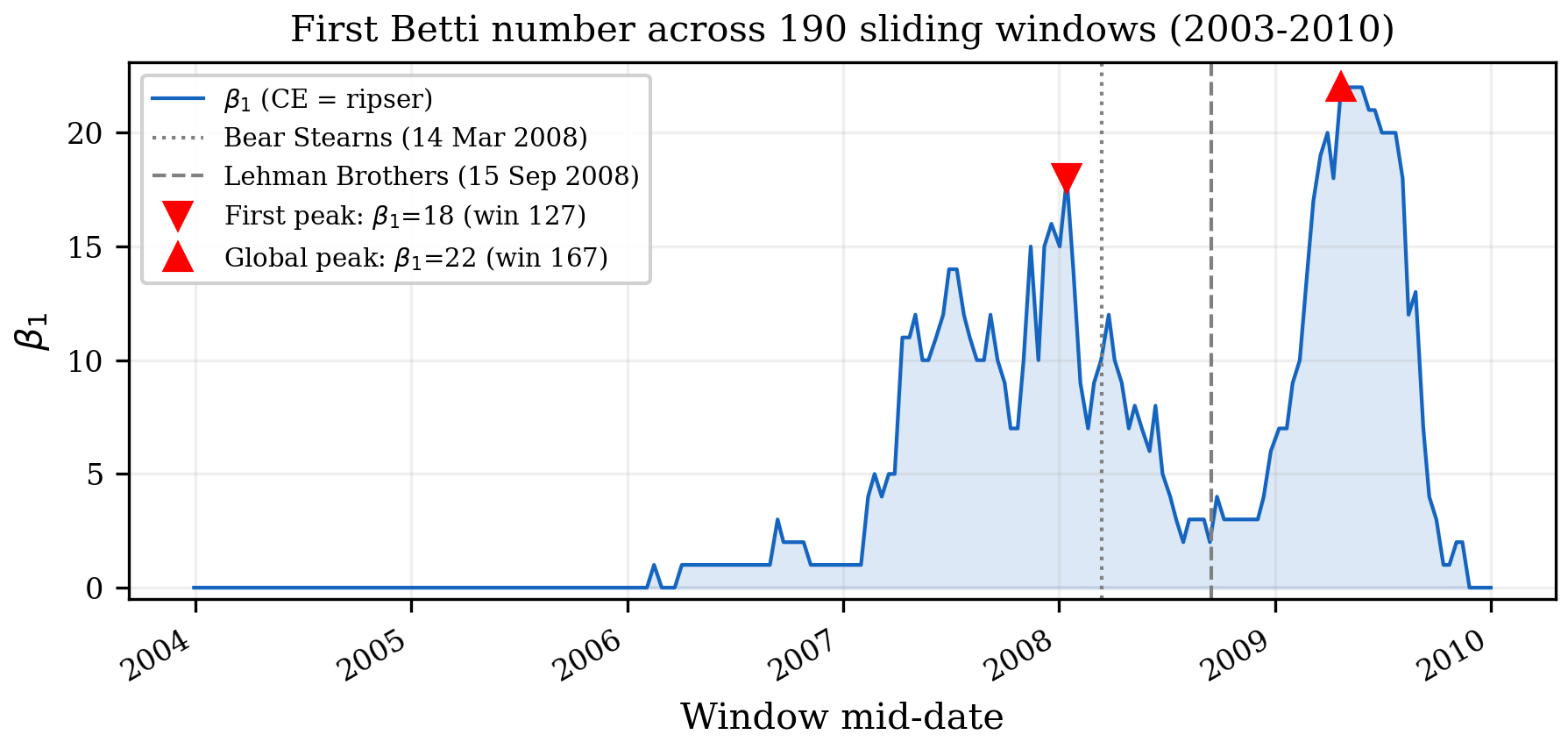}
  \caption{First Betti number $\beta_1$ over 190 sliding windows
    (2003--2010), computed by the classical CE benchmark and verified
    against \texttt{ripser}. Vertical lines mark Bear Stearns
    (14~Mar.\ 2008, dotted) and Lehman Brothers (15~Sep.\ 2008,
    dashed). $\beta_1 \in [0, 22]$; mean~$4.72$.}
  \label{fig:beta1_main}
\end{figure}

The first-Betti-number time series $\beta_1(t)$ across all 190 sliding
windows, computed by the CE benchmark, is shown in
Fig.~\ref{fig:beta1_main}. CE agrees with \texttt{ripser} on 190/190
windows (100\%) for both $\beta_0$ and $\beta_1$, confirming that the
classical stage of the pipeline is correct. The $\beta_1$
signal rises from zero in late 2006, reaches a first peak at
window~127 ($\beta_1 = 18$, January~2007 to January~2009), dips
briefly, and then attains the global maximum $\beta_1 = 22$ at
window~167 (April~2008 to April~2010). The topological complexity
begins building more than a year before the Bear Stearns collapse
(March~2008) and Lehman Brothers bankruptcy (September~2008),
consistent with the hypothesis that topological loop formation
provides a leading indicator of systemic stress~\cite{GK18}.
We emphasize that the curves in this section are produced by
the classical CE benchmark; PCE-VQE results on seven of these windows
appear in Sec.~\ref{sec:realwindow}.

\subsection{PCE Convergence Validation (Toy Laplacians)}\label{sec:pce_convergence}

\begin{figure}[t]
  \centering
  \includegraphics[width=\columnwidth]{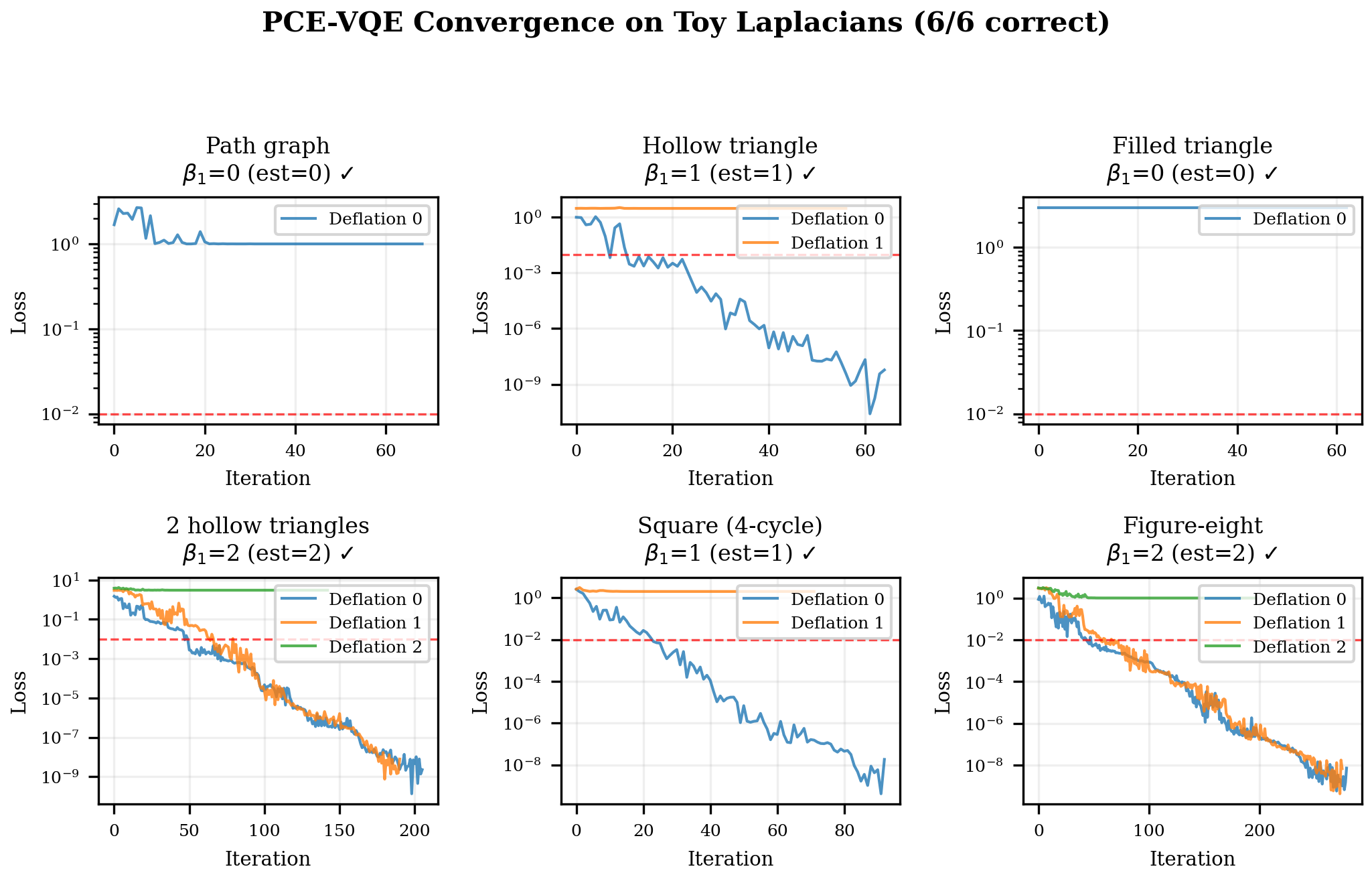}
  \caption{PCE-VQE convergence on six toy Laplacians (6/6 correct).
    Each panel is one toy Laplacian, labelled with its true $\beta_1$
    and the recovered estimate. Within a panel, each colored curve is
    one deflation round (round~0 blue, round~1 orange, round~2 green),
    that is, one search for a null vector orthogonal to those already
    found; the red dashed line is the null-space threshold
    $\delta=0.01$. A curve settling near $10^{-10}$ has located a null
    vector, while a round whose best loss jumps to ${\approx}3.0$ has
    none left to find and terminates the count. Each panel carries its
    own legend listing the deflation rounds it contains.}
  \label{fig:pce_toy_convergence}
\end{figure}

We validate the PCE-VQE pipeline on six synthetic Laplacians with
analytically known Betti numbers: path graph ($\beta_1{=}0$), hollow
triangle ($\beta_1{=}1$), filled triangle ($\beta_1{=}0$), two
disjoint hollow triangles ($\beta_1{=}2$), square/4-cycle
($\beta_1{=}1$), and figure-eight ($\beta_1{=}2$). All six are
recovered correctly with $\kappa{=}2$, $\delta{=}0.01$, $\mu{=}5.0$
(6/6 correct). Loss converges to ${\approx}10^{-10}$ for null vectors
and jumps to ${\approx}3.0$ when no further null vector exists
(Fig.~\ref{fig:pce_toy_convergence}).

The gradient-free deflation protocol correctly recovers
$\beta_1 \in \{1,2,3\}$ on synthetic Laplacians ($3/3$), but at
$\beta_1=4$ recovery becomes parameter-sensitive. This is the same
random-start landscape wall we isolate in Sec.~\ref{sec:realwindow}:
from a random start even an exact-gradient deflation finds no null
vector at $\beta_1=4$, yet with a classical-null-space warm start the
full $\beta_1=4$ and $\beta_1=6$ null spaces are recovered to machine
precision. The sensitivity
is therefore a symptom of the hardware-efficient-ansatz landscape, not
of penalty tuning, and is removed by the warm-started optimizer.

\paragraph*{Toy-to-real gap.} The validated PCE-VQE regime
($n_k \leq 64$, $\beta_1 \leq 4$) is roughly an order of magnitude
smaller in both dimension and null-space multiplicity than the real
S\&P~500 regime ($n_k\in[31,429]$; $\beta_1 \leq 22$). We probe this
gap directly in Sec.~\ref{sec:realwindow}, running PCE-VQE on the real
market Laplacians rather than on toy proxies of them.

\subsection{Real-Window Validation: PCE-VQE on Market Laplacians}
\label{sec:realwindow}

The toy Laplacians of Sec.~\ref{sec:pce_convergence} sit an order of
magnitude below the real-data regime. Since the 190 windows already
span every first Betti number from $0$ to $22$, we test transfer
directly: for each $\beta_1$ we take the smallest-$n_k$ real window
with that value (Table~\ref{tab:realwindow}), build $\Delta_1$ with the
\texttt{ripser}-verified construction of Sec.~\ref{sec:ce}, and run
PCE-VQE with the toy-validated configuration ($\kappa=2$, HEA depth
$L=2n$, COBYLA, $\delta=10^{-2}$). We report round~0, the search for
the first null vector; if its minimum loss stays above $\delta$ then
$\hat\beta_1=0$ regardless of later deflation. All seven windows use
$n\le 9$ qubits, so each runs in under two minutes on a CPU.

\begin{table}[t]
\caption{PCE-VQE on the real S\&P~500 window Laplacians (smallest-$n_k$
window per $\beta_1$; $\kappa=2$, $\delta=10^{-2}$).
$\min\mathcal{L}_0$ is the minimum round-0 loss from a random start
(HEA $L=2n$, COBYLA), which never reaches $\delta$, so $\hat\beta_1=0$.
$\min\mathcal{L}_{\Delta_1}$ is the minimum loss after warm-starting
from the classical null-space surrogate (exact-gradient L-BFGS, depth
$6n$; one extra qubit for window~67), which reaches machine zero and
recovers $\beta_1$.}
\label{tab:realwindow}
\centering
\begin{tabular}{rrrrrr}
\toprule
Window & $\beta_1$ & $n_k$ & $n$ & $\min\mathcal{L}_0$
       & $\min\mathcal{L}_{\Delta_1}$ \\
\midrule
67  & 1  & 76  & 5 & 1.29 & $1.4\times10^{-13}$ \\
185 & 2  & 105 & 6 & 1.27 & $1.5\times10^{-12}$ \\
85  & 3  & 99  & 6 & 1.21 & $9.2\times10^{-13}$ \\
98  & 4  & 160 & 7 & 1.74 & $4.5\times10^{-13}$ \\
157 & 6  & 167 & 7 & 1.47 & $4.8\times10^{-13}$ \\
117 & 10 & 176 & 7 & 1.80 & $4.9\times10^{-13}$ \\
170 & 22 & 259 & 9 & 1.48 & $4.8\times10^{-13}$ \\
\bottomrule
\end{tabular}
\end{table}

\paragraph*{Random-start failure.} PCE-VQE recovers no null vector on any window: the
minimum round-0 loss stays in $[1.21, 1.80]$ across $\beta_1=1$ to
$22$ (Table~\ref{tab:realwindow}), so $\hat\beta_1=0$ everywhere,
breaking already at $\beta_1=1$. The encoding is not the cause.
Optimizing over a fully expressive $n$-qubit state (all $2^n$
amplitudes free) drives the correlator vector onto the null space, with
overlap $1.00$ and Rayleigh quotient $0.00$ at $\beta_1=2$ (window~185)
and overlap $0.99$ at $\beta_1=1$ (window~67, exactly representable with
one extra qubit). The null vectors lie inside the reachable set; what
fails from a random start is the optimization. The tiny-gap $\Delta_1$
Rayleigh landscape traps COBYLA near $\mathcal{L}\approx 1.3$, and even
exact-gradient Adam and L-BFGS plateau near
$\mathcal{L}\approx 0.02$--$0.05$.

\paragraph*{Recovery by warm-starting.} This points to a fix. Given a
classical approximate null basis $N$ of $\Delta_1$, the surrogate
$B = I - N N^\top$ shares its null space but has spectral gap $1$, so
its Rayleigh quotient is benign to optimize. We minimize the PCE
Rayleigh quotient of $B$ with exact adjoint gradients and L-BFGS, then
polish on the true $\Delta_1$, using a depth-$6n$ ansatz and one extra
qubit where the minimal register cannot represent the null vector
exactly. This recovers a null vector on every window, driving the
$\Delta_1$ loss to ${\sim}10^{-13}$ across $\beta_1=1$ to $22$
(Table~\ref{tab:realwindow}, last column). The warm start uses the
classical null space, so this is a classical-quantum hybrid, not an
independent quantum determination of $\beta_1$: it shows the null
vectors are representable and reachable once the optimizer reaches the
right basin, so the random-start failure comes from the optimization
landscape. A quantum-independent pipeline would need
a quantum route to that warm start, such as a problem-informed or
adaptive ansatz (Sec.~\ref{sec:conclusion}).

\subsection{Encoding-Order Comparison
($\kappa=2$ vs.\ $\kappa=3$)}\label{sec:kappa_comparison}

On a benchmark Laplacian ($n_k = 64$, $\beta_1 = 2$; two disjoint
$32$-cycles), cubic PCE ($\kappa=3$, $n=5$) against quadratic
($\kappa=2$, $n=8$) cuts the qubit count by 37.5\% and the depth proxy
$d_{\mathrm{proxy}}=2n^2$ by 60.9\% ($128\to 50$), with both satisfying
$|\Pi^{(\kappa)}|\geq n_k$. With the exact-gradient, warm-started
optimizer of Sec.~\ref{sec:realwindow}, both encodings converge,
recovering $\beta_1=2$ to machine precision ($\mathcal{L}\approx
5\times10^{-15}$ for $\kappa=2$, $2\times10^{-11}$ for $\kappa=3$);
a gradient-free optimizer at fixed budget left both short of $\delta$.
So the cubic encoding attains the same recovery at strictly lower qubit
count and depth.

\subsection{Resource Scaling Analysis}

Table~\ref{tab:resources} presents analytical resource estimates
comparing PCE ($\kappa{=}2$) with LGZ-QPE across a range of Laplacian
dimensions.

\begin{table}[t]
\caption{Resource Estimates: PCE ($\kappa{=}2$) vs.\ LGZ-QPE}
\label{tab:resources}
\centering
\begin{tabular}{rcccc}
\toprule
$n_k$ & PCE qubits & PCE depth & LGZ total qubits & LGZ depth \\
\midrule
64      &   8  &     128  & 13 & $>10^4$ \\
256     &  16  &     512  & 15 & $>10^4$ \\
1\,024  &  32  &   2\,048 & 17 & $>10^5$ \\
4\,096  &  64  &   8\,192 & 19 & $>10^5$ \\
16\,384 & 128  &  32\,768 & 21 & $>10^6$ \\
65\,536 & 256  & 131\,072 & 23 & $>10^6$ \\
\bottomrule
\end{tabular}
\par\smallskip
{\footnotesize Analytical estimates. PCE depth $= 2n^2$; LGZ depth
$= O(N_P \cdot r \cdot 2^p)$ for a Laplacian with $N_P$ Pauli terms,
$r$ Trotter steps per controlled evolution, and
$p = \lceil\log_2(1/\epsilon)\rceil$ ancilla qubits (not shown). PCE uses more qubits but far shallower
circuits and no ancilla overhead.}
\end{table}

Table~\ref{tab:resources} also contextualizes the depth--qubit
trade-off against other quantum TDA methods. LGZ-QPE and the
streamlined algorithm of~\cite{McArdle22} use only $O(\log n_k)$
qubits but require deep circuits ($O(N_P \cdot r \cdot 2^p)$ for
standard QPE). Thermal QTDA~\cite{Scali24} matches the logarithmic
qubit scaling with $2\lceil\log_2 n_k\rceil$ system qubits and a
single ancilla, with depth governed by the Lindbladian mixing time.
PCE uses more qubits ($O(n_k^{1/\kappa})$) but eliminates ancilla
overhead entirely and operates at circuit depths polynomial in
$n_k^{1/\kappa}$, which matters on near-term hardware where circuit
depth is the binding constraint.

\subsection{Noise Robustness}\label{sec:noise_results}

We compare PCE-VQE against LGZ-QPE under depolarizing noise at error rates $p \in \{0,\, 10^{-4},\, 5{\times}10^{-4}, \, 10^{-3},\,
5{\times}10^{-3},\, 10^{-2}\}$. PCE-VQE: density-matrix simulation
at $n=6$ qubits on 4 test cases ($\beta_1 \in \{0,1,2,3\}$, 5 trials
each). LGZ-QPE: matched-scale simulation at $n_k\in\{8,16\}$ plus an
analytical extrapolation based on worst-case depolarizing
accumulation over the QPE depth.

\paragraph*{Comparison at $n_k=16$.} At matched $n_k = 16$, the LGZ-QPE simulated accuracy
drops from 1.00 at $p=0$ to 0.316 at $p=10^{-4}$ and below $10^{-2}$
at $p=10^{-3}$, while PCE-VQE maintains accuracy above 0.90 through
$p=5\times 10^{-3}$. This comparison is not at matched
encoded problem size: PCE is at fixed $n=6$, encoding $n_k\approx
63$ simplex dimensions under $\kappa=2$, while LGZ is at fixed
$n_k=16$. The figure therefore reflects the depth difference between
the two encodings at moderate scale, not a like-for-like
accuracy-vs-noise comparison. It supports only the qualitative
conclusion that the shallower PCE-VQE circuits accumulate less
depolarizing error per shot than depth-heavy LGZ-QPE; it does not
quantify an advantage in tolerable per-gate error at matched task
accuracy.

\subsection{Barren Plateau Analysis}\label{sec:bp_results}

\begin{figure}[t]
  \centering
  \includegraphics[width=\columnwidth]{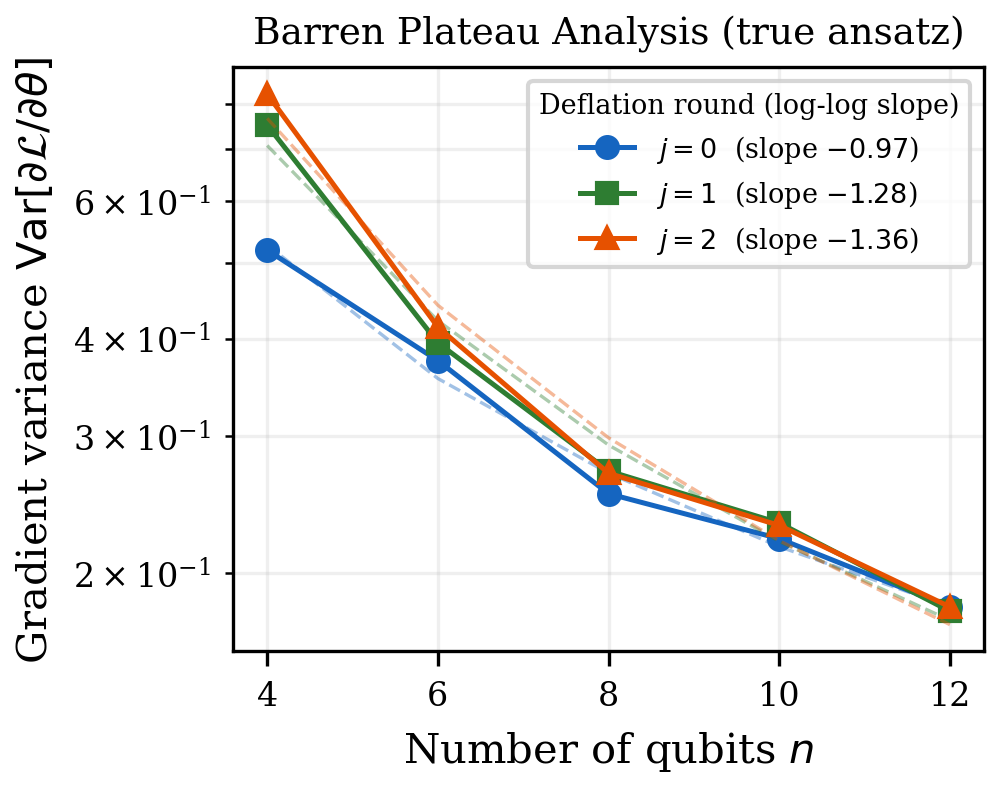}
  \caption{Gradient-variance scaling of the true PCE Rayleigh-quotient
  loss through the hardware-efficient ansatz (exact adjoint gradient;
  deflation rounds $j\in\{0,1,2\}$, $n=4$--$12$). The variance decays
  only polynomially (log--log slopes ${\approx}{-}1$), staying order
  $0.2$ at $n=12$: a mild decrease, not the exponential vanishing of a
  barren plateau.}
  \label{fig:barren_plateau}
\end{figure}

We probe trainability through the gradient variance
$\mathrm{Var}[\partial\mathcal{L}/\partial\theta_\ell]$ of the true PCE
Rayleigh-quotient loss, evaluated exactly through the
hardware-efficient ansatz with the adjoint gradient of
Sec.~\ref{sec:realwindow} over 100 random initializations, taken as a
median over sparse random combinatorial Laplacians
(Fig.~\ref{fig:barren_plateau}). Over $n=4$ to $12$ the variance decays
only \emph{polynomially}: the log--log slopes of variance vs.\ $n$ are
$-0.97$ ($j{=}0$), $-1.28$ ($j{=}1$), and $-1.36$ ($j{=}2$), with the
variance still of order $0.2$ at $n=12$. This mild, roughly $1/n$
decrease is far from the exponential vanishing that defines a barren
plateau and is consistent with the super-polynomial variance lower
bound PCE enjoys for its bilinear loss~\cite{Sciorilli25}. That bound
is not proved for our rational deflation loss
(Sec.~\ref{sec:deflation}), so this is empirical evidence of
trainability, not verification of the asymptotic bound; we make no
claim beyond $n=12$.

\subsection{Classification Performance and Generalization Failure}
\label{sec:classification}

We evaluate whether the $\beta_1$ signal provides predictive value
for crash-risk screening using a chronological train/test split. A
crash event is defined as a drawdown exceeding 10\% within 90
calendar days; under this definition, 37 of the 190 windows are
positive. The dataset is split chronologically: windows whose start
dates fall in 2003--2006 form the training set (used to select the
classifier threshold), and windows starting in 2007--2010 form the
held-out test set. Threshold selection maximizes $F_1$ on the
training split only. The volatility-proxy baseline, the 30-day
rolling standard deviation of log returns, is evaluated under the
identical protocol.

On the held-out test set, the $\beta_1$-threshold classifier achieves
precision~0.278, recall~1.000, $F_1$~0.435, and ROC AUC~0.818, compared
to ROC AUC~0.652 for the volatility proxy
(Table~\ref{tab:classification}). We caution that the test split
contains only 10 positive windows (of 64) over 4 years; the recall of
$1.000$ at precision $0.278$ means the classifier fires on a large
fraction of test windows rather than acting as a sharp early-warning
signal. A bootstrap 95\% confidence interval on the test ROC AUC
(resampling test windows, 1000 resamples) spans $[0.71, 0.91]$, wide
enough that the point estimate should not be over-read. To test
significance despite the small sample, we permute the held-out labels
$20{,}000$ times and recompute the test AUC: the observed $\beta_1$ AUC
of $0.818$ is reached by only a $p=6\times10^{-4}$ fraction of
permutations, so it is well beyond chance, whereas the volatility proxy
($0.652$, $p=0.07$) is not significant at the $5\%$ level.
Both the bootstrap and the permutation test treat windows as
independent; consecutive windows share $492$ of $500$ days, so the
effective sample size is much smaller than 64 and these intervals and
$p$-values are optimistic. In-sample
metrics on the training split are reported for reference. For robustness, we additionally evaluate on two
independent out-of-distribution crisis episodes not used for training
or threshold selection: the 2020 COVID shock and the 2022 rate-cycle
drawdown, with the fixed 2003--2006-trained threshold
(Table~\ref{tab:classification_oos}).

\begin{table}[t]
\caption{Classification Performance: Chronological Split}
\label{tab:classification}
\centering
\begin{tabular}{lcccc}
\toprule
Classifier / split & Precision & Recall & $F_1$ & ROC AUC \\
\midrule
$\beta_1$ (train, 2003--2006) & 0.870 & 0.741 & 0.800 & 0.943 \\
$\beta_1$ (test,  2007--2010) & 0.278 & 1.000 & 0.435 & 0.818 \\
Volatility (train)            & 0.545 & 0.889 & 0.676 & 0.905 \\
Volatility (test)             & 0.146 & 0.700 & 0.241 & 0.652 \\
\bottomrule
\end{tabular}
\par\smallskip
{\footnotesize Test ROC AUC for $\beta_1$: bootstrap 95\% CI
$[0.71, 0.91]$ (1000 resamples); label-permutation $p=6\times10^{-4}$.
Test set has 10 positive windows of 64; metrics should be interpreted
with attention to small-sample variance.}
\end{table}

\begin{table}[t]
\caption{Out-of-Distribution Evaluation (Fixed 2003--2006 Threshold)}
\label{tab:classification_oos}
\centering
\begin{tabular}{lcccc}
\toprule
Episode & Precision & Recall & $F_1$ & ROC AUC \\
\midrule
2020 COVID shock          & 0.000 & 0.000 & 0.000 & 0.009 \\
2022 rate-cycle drawdown  & 0.333 & 0.083 & 0.133 & 0.515 \\
\bottomrule
\end{tabular}
\end{table}

\begin{figure}[t]
  \centering
  \includegraphics[width=\columnwidth]{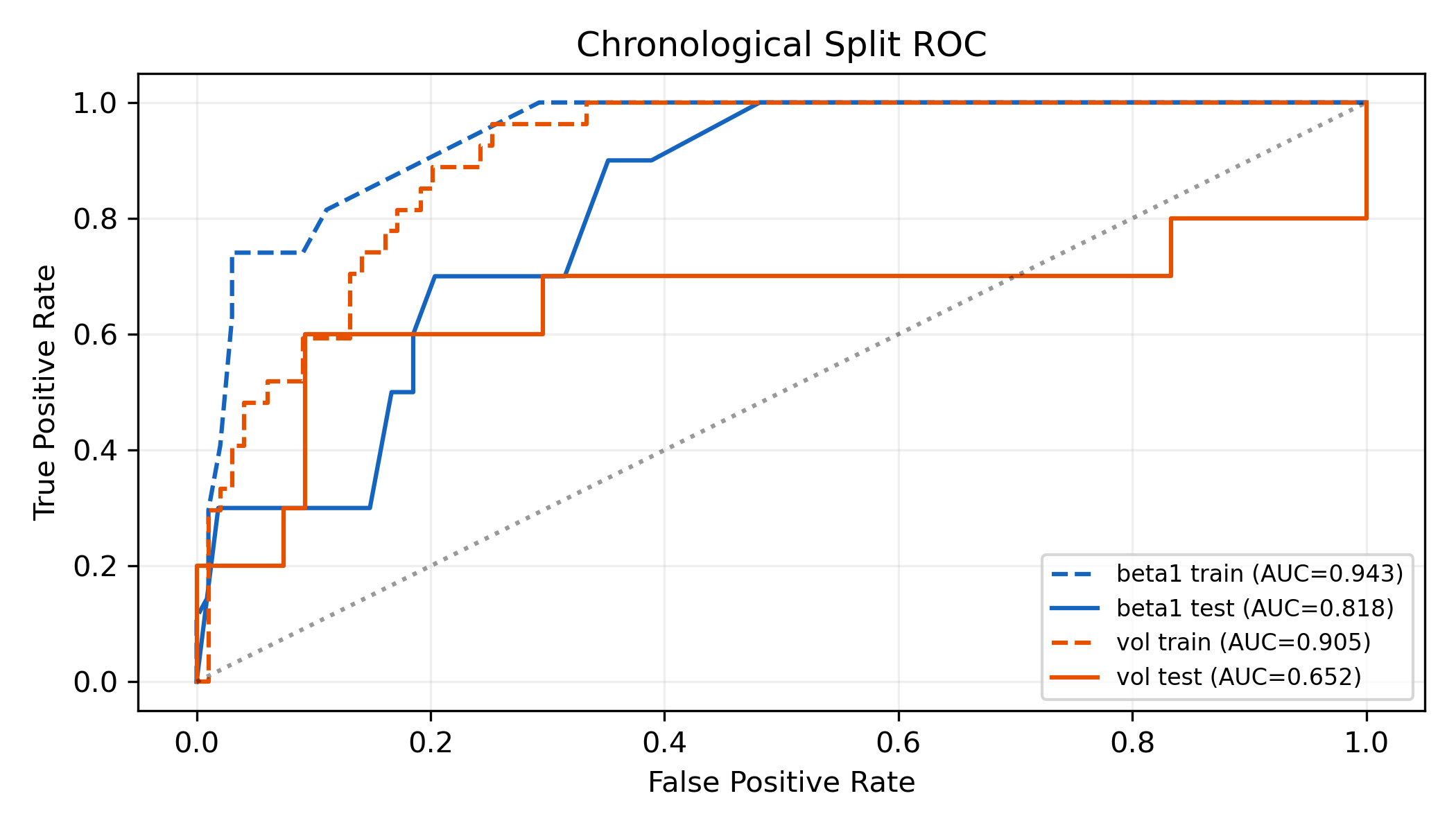}
  \caption{ROC curves for the $\beta_1$ classifier and volatility
  proxy under the chronological train/test split. Training curves
  (dashed) show in-sample performance on 2003--2006; test curves
  (solid) show held-out performance on 2007--2010.}
  \label{fig:roc_classification}
\end{figure}

\paragraph*{Out-of-distribution failure.} A test ROC AUC of $0.009$ on the COVID episode
is essentially the \emph{inverted} classifier. Because ROC AUC does
not depend on the decision threshold, this reflects the $\beta_1$
ranking itself: during the COVID shock, $\beta_1$ is
\emph{anti-correlated} with crash labels. Two readings are consistent:
a mechanism mismatch (the 2008 GFC was a slow build-up of correlated
stress that loop-formation tracks, whereas the 2020 shock was a fast
exogenous dislocation with different return-series topology), or a
scale artifact (the filtration scale $\varepsilon^*$ and embedding
parameters were fixed on GFC-era data and may not transfer to a
different volatility regime). The 2022 chance-level
AUC of $0.515$ fits either, and disentangling them needs a
regime-conditional classifier, left to future work. The in-regime AUC
of $0.818$ is thus evidence that $\beta_1$ carries structural
information about GFC-style correlated-stress build-up, not that a
single fixed-threshold $\beta_1$ classifier is a universal
early-warning indicator.

One clarification on scope: this classification study uses the
classical exact-eigensolver $\beta_1$ of Sec.~\ref{sec:ce}, not a
quantum estimate. The role of the quantum pipeline is to compute that
same topological invariant, so the finding here is a property of the
$\beta_1$ feature and its regime dependence, and it holds whether
$\beta_1$ is obtained classically or by PCE-VQE. It is therefore a
statement about the topological signal, not a claim about quantum
learning.

\subsection{Near-Term Hardware and Cost Crossover (Estimates)}
\label{sec:hardware_estimate}

To gauge executability, we estimate the cost of PCE-VQE at $n_k = 256$
($n = 16$, $\kappa=2$) on an IBM Heron-class device
($\epsilon_{2Q}=3\times10^{-3}$, $\epsilon_{1Q}=2\times10^{-4}$).
Transpiled to $\{CX, SX, R_Z\}$, the $L=32$ ansatz has ${\approx}480$
two-qubit and $2640$ single-qubit gates, for a per-shot fidelity
$(1-\epsilon_{2Q})^{480}(1-\epsilon_{1Q})^{2640}\approx 0.14$. At
$10^4$ shots per basis, $G=3$ bases, $200$ iterations per round, and
$\beta_1=2$ rounds, the shot budget is ${\approx}1.2\times10^7$, within
the throughput of current cloud devices; scaling to $n_k\geq 10^3$
would need error mitigation or early fault-tolerant hardware. We have
not run this on hardware. Extrapolating the same model against the CE
benchmark (all 190 windows in $\approx 25.5$~s) puts the cost crossover
near $n_k\gtrsim 10^4$, where $O(n_k^3)$ diagonalization overtakes the
projected per-evaluation cost. That threshold compounds several
extrapolations and is a research target, not a demonstrated crossover;
datasets that reach it (multi-asset networks, tick-level embeddings)
are plausible.

\section{Discussion}\label{sec:discussion}

\paragraph*{Structural advantages.} The PCE-variational approach
offers no ancilla overhead (vs.\ $O(\log(1/\epsilon))$ for QPE) and
shallow $O(n^2)$ circuits whose gradient variance decays only
polynomially (not exponentially) across $n=4$--$12$, consistent with
PCE trainability~\cite{Sciorilli25} even though our rational loss is
not formally covered by that proof. The matched-scale noise simulations at
$n_k \in \{8, 16\}$ show PCE-VQE holding $>$90\% accuracy where LGZ-QPE
collapses; as Sec.~\ref{sec:noise_results} notes, this illustrates the
depth/noise trade-off rather than a calibrated matched-encoding
benchmark.

\paragraph*{Limitations.} For $\kappa=2$ PCE uses $O(\sqrt{n_k})$
qubits vs.\ $O(\log n_k)$ for QPE, so the advantage is in depth, not
qubit count. The variational deflation protocol has no performance
guarantee and spends one optimization per null vector, so its total
budget grows linearly in $\beta_k$ (the $\beta_1=4$ sensitivity of
Sec.~\ref{sec:pce_convergence} is a symptom of the random-start
landscape rather than a separate issue). This points to
subspace methods such as SSVQE or weighted subspace search that target
several eigenvectors at once, which we expect to be necessary for
higher $\beta_k$. We also make no universal-approximation claim for
the encoding. The null-space task needs something weaker than full
expressivity, namely that the reachable set $\{\mathbf{c}(\vec\theta)\}$
intersect $\ker(\Delta_k)$, and for deflation contain a spanning set of
it, rather than cover all of $\mathbb{R}^{n_k}$. The real-window
experiment of Sec.~\ref{sec:realwindow} separates the failure modes.
For a fully expressive state the $\kappa=2$ correlators reach the null
vectors of the real Laplacians (overlap $\approx 1$, with a one-qubit
margin at the minimal register size) even at $\beta_1=22$, so the
encoding satisfies the feasibility condition. From a random start the
hardware-efficient ansatz nonetheless recovers no null vector, but
warm-starting from a classical null-space surrogate recovers $\beta_1$
exactly on all seven windows tested. A fully quantum recovery therefore needs a
quantum route to that warm start.

\paragraph*{Possible regime of use.} If the toy-to-real gap
is closed, PCE-variational would suit near-term hardware where depth is
the binding constraint and moderate-scale problems
($n_k\sim 10^3$--$10^5$) where classical $O(n_k^3)$ diagonalization is
slow but full fault-tolerant QPE is not yet justified. We claim no
quantum advantage at the scales demonstrated: at $n_k\in[31,429]$
classical methods solve each window in $\leq 228$~ms, so the
contribution is methodological.

\section{Conclusion}\label{sec:conclusion}

We presented, to our knowledge, the first application of Pauli
Correlation Encoding to quantum topological data analysis: a
depth-efficient pipeline that recasts Betti-number estimation as a
continuous-PCE Rayleigh-quotient minimization with deflation over
shallow, ancilla-free circuits, at the cost of a rational loss outside
the trainability proof of~\cite{Sciorilli25}, for which the gradient
variance decays only polynomially (no exponential barren plateau)
over $n=4$--$12$ qubits. Numerical
experiments on S\&P~500 data validate the classical stage against
\texttt{ripser}. Run end-to-end on the
real market Laplacians ($\beta_1=1$--$22$), the quantum stage recovers
no null vector from a random start, but warm-starting from a classical
null-space surrogate recovers $\beta_1$ exactly on all seven windows
tested, up to $n_k=259$ (Sec.~\ref{sec:realwindow}). The encoding contains the null vectors,
so the obstacle is the optimization landscape of the
hardware-efficient ansatz. Chronologically split
classification gives in-regime ROC AUC $\approx 0.818$, but the
calibration fails to generalize to the 2020 COVID and 2022 rate-cycle
regimes, indicating that any deployment would require regime-aware
calibration. Resource estimates extrapolate a crossover at
$n_k\gtrsim 10^4$. The principal
open question is therefore whether a quantum-native warm start, from a
problem-informed or adaptive ansatz, can replace the classical
null-space guidance used here and recover $\beta_1$ independently at
real-data scale.

Future work, ordered by priority: (i)~a quantum-native warm start, via
problem-informed or adaptive ans\"atze, so that the real-window
recovery of Sec.~\ref{sec:realwindow} becomes independent rather than a
classical-quantum hybrid; (ii)~subspace deflation (SSVQE, weighted
subspace search) for higher-multiplicity null spaces; (iii)~regime-aware
calibration of the $\beta_1$ classifier; (iv)~persistent Betti numbers
via~\cite{Hayakawa22}; (v)~hardware execution
following~\cite{Sciorilli25, Sciorilli25LABS}; and (vi)~formal
expressivity and trainability statements for the continuous-PCE loss.

\section*{Data Availability}

The S\&P~500 dataset, pipeline code, and PCE-VQE simulation scripts
are available at \url{https://github.com/arulrhikm/Quantum-Market-Crash-TDA}.

\section*{AI Use Disclosure}

In accordance with IEEE policy, we disclose that large language models
were used only for writing assistance (clarifying prose and phrasing).
All technical content, results, code, and figures were produced and
verified by the authors; no AI system generated experimental data or
numerical results. The authors take full responsibility for the
accuracy, originality, and integrity of the submission, including every
cited reference.

\section*{Acknowledgment}

Classical computations were performed on Google Colab. Quantum
statevector simulations were performed on BlueQubit.

\bibliographystyle{IEEEtran}
\bibliography{references}

\end{document}